\documentstyle[12pt]{article}
\begin{document}
\begin{center} {\bf A COMPUTER SIMULATION STUDY OF IONIC CONDUCTIVITY IN
POLYMER ELECTROLYTES} \end{center}
\begin{center} Aninda Jiban Bhattacharyya\footnote[1]{
e-mail:aninda@juphys.ernet.in}, T.R.Middya and S.Tarafdar
\newline  Condensed Matter  Physics Research Center, Department of
Physics, \newline 
Jadavpur University, Calcutta-700032, INDIA
\end{center}
\begin{flushleft}
\bf Pacs No. : 66.30.Dn; 61.43.Bn; 61.41.+e\newline
\end{flushleft}
\begin{abstract}
In this paper we present a computer simulation study of ionic
conductivity in solid polymeric electrolytes. The multiphase nature of
the material is taken into account. The polymer is represented by a
regular lattice whose sites  represent either crystalline or amorphous
regions with the charge carrier performing a random walk.
Different waiting times are assigned to sites corresponding to the different 
phases. A random walk (RW) is used to calculate the
conductivity through the Nernst-Einstein relation. Our walk algorithm takes 
into account the reorganisation of the
different phases  over time scales comparable to time scales for the
conduction process. This is a characteristic feature of the polymer
network. The qualitative nature of the variation of conductivity with
salt concentration agrees with the experimental values for
 PEO-NH$_{4}$I and
PEO-NH$_{4}$SCN. The average jump distance estimated from our work is
consistent with the reported bond lengths for such polymers. 
\end{abstract}
\newpage
\section {Introduction }

During  the last two decades a lot of interest has been generated in
the potential
industrial applications of  polymer electrolytes [1,2]. Hence, 
developement  of
theoretical understanding of such materials is essential.
Polymeric solid electrolytes are  formed by
complexing an ionic salt like NaI, NH$_{4}$SCN, etc with polymers, for
example polyethylene oxide (PEO), polypropylene oxide (PPO), [3,4] etc. We
restrict our discussion to solvent free polymer electrolytes 
which are formed
by dissolving or suspending the salt and the polymer in a suitable
 solvent and
then evaporating the solvent  during casting. 

The theoretical explanation of the dependence of ionic transport on
temperature, salt fraction, frequency, etc. is a formidable problem. The
complications arise mainly because  these polymer systems are
multiphase  in nature. Further difficulties creep in because the presence
of different  phases are dependent on processing conditions and
thermal history.  

The aim of this work is to study the variation of ionic conductivity in a
polymer electrolyte where the salt fraction is varied. The varying salt
fraction changes the proportion of crystalline and amorphous regions and
also the concentration of charge carriers supplied by the
salt [5]. 

The final goal is of course to be able to predict the
conductivity of an unknown polymer complex with a given salt 
fraction. In this 
work we have not fully achieved this since we have to use the experimental 
data for crystallinity at different salt fractions as input. We
first briefly describe previous work done in this field and then present
our model.
 
Since analytical calculation of conductivity in such a complex system is
very difficult, a class of models frequently used to study conductivity in
disordered media involve computer simulation of a random walk (RW) [6-8].
Appropriate algorithms to represent phases of different conductivity may
be incorporated into such models. Often a continuous time random
walk (CTRW) [8] is used. A recent work [9] introduces a method
 for speeding up
a CTRW calculation by  a modified blind ant algorithm which makes the
walker jump at every step but adjusts the time to account for the waiting
time. 

The dynamic bond percolation model by Druger et al [10,11] is used to
calculate the conductivity of a polymer through computer simulation of a
hopping model.
This model incorporates the dynamic nature of the medium by introducing a
renewal time. The frequency dependence of the conductivity within a certain
range can be explained by this model. The model fails however at the very
high frequency limit, since inertial effects are not taken into account.
\section{\bf The Model } 

In the present paper our main aim is to develop an algorithm utilising
some experimental inputs to  explain the salt fraction dependence of ionic
conductivity of polymer-salt systems.
The solvent free polymer electrolytes are prepared in the form of thin
films, so as a preliminary approximation, in this paper we consider the
polymer electrolyte as a two dimensional space lattice. Rather than
 dealing with
bonds, our model deals with sites only. A random walk is executed on the
square lattice. The uniform square lattice does not of course represent
the structure of the polymer network which is disordered, but only
provides a convenient space for executing the random walk.

We designate a site on the lattice as amorphous or crystalline. 
The  sites actually represent  very small regions of the polymer complex
which belong to a single phase only, and the lattice spacing ($\xi$)
represents the distance between such sites.
We assign a jump probability to each site representing the
conductivity of the phase to which the site belongs. 

However, our model is not a "quenched" model. The walker sees a site as
crystalline or amorphous statistically according to a random number
chosen on its visit to the site. The probability of the site being,
 say crystalline, is determined
from the experimental crystallinity. This means that a site previously
found to be crystalline may become amorphous on a later visit. A
crystalline site may also become amorphous or vice versa, during the
waiting  time of the walker at that site.

This procedure is often
adopted to make the walk algorithm simpler and provide an effectively
infinite system. Here however, it gives an added advantage. The
 polymer, as
distinct from crystalline or glassy superionic conductors is  known to
show a dynamic disorder, i.e. the matrix is continually rearranging itself
after  a characteristic (renewal) time  $\tau_{r}$. The RW algorithm thus
gives a more realistic picture of the polymer, though in the present
formalism we have no control over $\tau_{r}$. We are at present working in
a variation of the model where $\tau_{r}$ can be varied, this is to be
reported shortly [12].

\subsection{\bf Random Walk Algorithm } 
The  polymer-electrolyte system is
multiphase in nature consisting of both crystalline (belonging 
to complexed and
uncomplexed polymer) [1,3,4] and amorphous regions. For simulation at a
particular salt fraction, the ratio of amorphous to crystalline sites is 
taken from experimental estimates of crystallinity from differential
thermal analysis (DTA) or x-ray diffraction (XRD), and the
distribution of these phases in the lattice is random. A site
 belonging to the $i^{th}$
phase is assigned a jump probability $p_{i}$ for jumping to a
 nearest neighbour
site, at each time step. The time elapsed (inverse of p$_{i}$) between
arrival at one site and arrival at the next site includes $\tau_{j}$
 the time 
taken to jump from one site to a nearest neighbour and an average waiting 
time <$\tau_{i}$> at the i$^{th}$ site. A longer waiting time corresponds 
to a lower conductivity of
the phase and hence of the site. In the present case we have
 two waiting times, $\tau_{a}$
for the amorphous phase and $\tau_{c}$ for the crystalline phase which is
larger. 

A distribution of energetically different sites on a lattice is usually
represented in a simulation model by either of the following pictures :
\newline  
(a) A well model, where each site is a potential well. Here the well
depth w$_{i}$ is characteristic of that particular site i, and
determines how long the random walker will be trapped there.\newline
(b) A barrier model where a barrier of height
h$_{ij}$, is envisaged between the sites i and j. Here the probability of
hopping to i from j may be  determined by the nature of both the 
sites i and j.

In the present work we employ the considerably simpler well model, where
the probability of leaving a site is determined by the phase of this site,
but the carrier has an equal probability of going to all four nearest
 neighbours
whether they are crystalline or amorphous. In view of the dynamic
disorder,  where the polymer chains can reorient within the time
 interval $\tau$,
which the random walker takes to hop to a neighbouring site,
 model (a) seems quite adequate.

In  case of normal diffusion, the diffusion coefficient can be
obtained from a random walk, through the relation 
\begin{equation}
<r^{2}(t)>=2dDt
\end{equation}
where $<r^{2}>$ is the average square distance covered by the walker in t
time steps and d is the dimension of the space lattice. The constant D is
the diffusion coefficient.  \newline 
In real units the diffusion coefficient is given by
\begin{equation}
D=\frac{<r^{2}>}{4t}\frac{(distance\>unit)^{2}}{(time\>unit)}
\end{equation} The distance unit = the lattice constant ($\xi$) and time
unit = $\tau$, the time step. In our model $\tau$ = $\tau_{j}$ the time
taken in jumping from one site to another, waiting time at a site is
measured in units of $\tau$. The steps of the RW are given below 
\begin{enumerate} \item The walker is released at a randomly chosen site
on a two dimensional
lattice. \item  A random number R$_{1}$ is chosen. If R$_{1}$ < c (c is
the crystallinity), the site is crystalline, otherwise it is amorphous.
\item  Now a second random number R$_{2}$ is chosen. The residence
probability of the appropriate site is (1-p$_{i}$). So the
 probability of jumping  to a
particular neighbour is p$_{i}$/4 \newline 
If \[ 0 \leq R_{2} \leq  \frac{p_{i}}{4} \] the walker jumps to
the left neighbour. \newline If \[  \frac{p_{i}}{4} <  R_{2}
\leq  \frac{p_{i}}{2} \] the walker jumps to the right neighbour. \newline
If \[ \frac{p_{i}}{2} <  R_{2} \leq 
\frac{3p_{i}}{4}  \] the walker moves to the upper neighbour
\newline and if \[ \frac{3p_{i}}{4} < R_{2} \leq p_{i} \] the walker moves
 to the lower neighbour.
 \newline If \[ p_{i} < R_{2} \leq 1 \] 
it does not jump at all. \end{enumerate} In case the walker
has jumped, it again checks whether the new site is crystalline or
amorphous by step(2). Even if it has not jumped step(2) is repeated to
account for the reorganisation of the lattice i.e. an amorphous site may
become crystalline after each time step. 

The walk proceeds in this manner for the requisite number of steps. Due to
the stochastic nature of the process, one has to average over a large
number of such walks to get a meaningful value of <r$^{2}$>. In this work
the walker does a random walk of (15000-75000) steps and distance (r)
covered is averaged over (20000-100000) walks. This gives sufficiently
good convergence (up to 3 significant figures) for the diffusion
coefficient.

Our random walk algorithm allows the walker to move  on an effectively
infinite sample. This is possible because here we do not take a quenched
system  with sites assigned specifically to a definite phase. So the
problem of finite size effects is avoided to some extent. But, of course
there is a limitation to the size of the walk due to
restricted computer time. For normal diffusion the relation between the
diffusion coefficient and the conductivity is given by the Nernst-Einstein
equation 
\begin{equation}
\sigma = \frac{DNq^{2}}{kT}
\end{equation}
where N is the number density of the mobile ion, q is the  charge of the
mobile ion, $k$ is the Boltzmann's constant and T is the temperature.
The above equation is written as 
\begin{equation}
\sigma=KXD 
\end{equation}
where K is a constant and  X is the salt fraction.
Since in eqn.(3) $k$, q and  T are constants for conductivity
 measured at different
salt fractions we have conveniently grouped them into a single constant
K.  In  ion conducting polymers, mobile charge carriers are provided by
the  salt, the pure polymer being an insulator. So N which is the number 
density of charge carriers is assumed to be proportional to
 the salt fraction X in eqn.(3).
\section {\bf Waiting times and jump distance in terms of the
simulation parameters } 

A random walk is usually performed with the time step and lattice constant
chosen to be unity for convenience. This will give D in terms of arbitrary
units. To compare the value obtained with an  experimental result, we
must assign real values to $\tau$ and $\xi$. This
section identifies these quantities in terms of experimentally measurable
properties. 

Our model lattice contains two types of sites,  crystalline (c) and
amorphous (a). The relative number of each depends on the salt
fraction. Suppose $\tau$ is the time unit for the walk. Let the
average waiting  time for the amorphous and crystalline sites
 be <$\tau_{a}$> and 
<$\tau_{c}$> respectively. Let $\tau_{j}$ be the jump time between sites.
Total time t for the walk is given by 
\begin{equation} t = N_{c}<\tau_{c}+\tau_{j}>+N_{a}<\tau_{a}+\tau_{j}>
\end{equation}   We assume here
$ \tau_{j} = \tau $ which is a constant.  So eqn.(5) becomes,
\begin{equation} t = N_{c}(t_{c}+1)\tau + N_{a}(t_{a}+1)\tau
\end{equation}
where $N_{c}$ and $N_{a}$ are respectively the total number of
 crystalline and
amorphous sites visited during the walk. t$_{c}$ and t$_{a}$ are the
waiting times measured in units of the jump time $\tau_{j}$ (= $\tau$). 
\newline Eqn (6) can be written as 
\begin{equation} t = \tau (t_{a}+1) \left[N_{a} + N_{c}
\frac{(t_{c}+1)}{(t_{a}+1)} \right]
\end{equation}
Now $(t_{c}+1)$ and $(t_{a}+1)$  are nothing but
inverse of the probability for a walker to jump from crystalline and
amorphous sites  respectively.
\begin{equation}
t_{c}+1=\frac{1}{p_{c}} : t_{a}+1=\frac{1}{p_{a}}
\end{equation}
where p$_{c}$ and p$_{a}$ are the  jumping probabilities
from crystalline and amorphous sites respectively. p$_{i}$ may vary from
0 to 1, corresponding to the lowest and highest possible
 conductivities. For
p$_{i}$ =0, t$_{i}$ is $\infty$, the carrier getting permanently
trapped and for p$_{i}$ =1, $t_{i} = 0$  \newline 
We now define 
\[\frac{t_{a}+1}{t_{c}+1} = \frac{p_{c}}{p_{a}}=\frac{1}{R}\]
A time interval of N$_{t}$ time steps corresponds to a real time
 interval of N$_{t}\tau$  seconds.
To calculate $\tau$, we use an estimate of $\tau(t_{a}+1)$, 
obtained from NMR linewidth narrowing measurements [13-15]. 
This is a measure of the average time interval between succesive
 jumps. Now we can 
use eqn.(2) to calculate $\xi$, by comparing (<r$^{2}>/4t$)
 from our simulation, 
with a typical experimental value of D [16]. This
 gives $\xi$ = 6.14 A which is physically 
reasonable being of the order of the interatomic spacing [2].  
We now calculate the variation of D and hence
 $\sigma$ with the salt fraction 
X. \section
{\bf  Results and discussion  } 

Using the above algorithm we have calculated the diffusion coefficients as
functions of salt fractions of two polymer electrolytes
 --- PEO-NH$_{4}$SCN and PEO-NH$_{4}$I. The proportion of
 crystalline and amorphous
sites for the simulation was obtained from experimental estimates of
crystallinity  from DTA measurements [3,4]. The parameter R 
is as given  in Table 1. We find R=9.99 gives
satisfactory results in both cases showing that it is a property of the
host polymer PEO. The conductivity is
estimated using eqn.(4). In eqn.(4) X are taken from experiment [3,4] and
K is an arbitrary constant  as given
in Table 1. The calculated values for conductivity fitted well with the
experimental values (at room temperature) as shown in Fig.1.
 and Fig.2. [3,4].
For a pure polymer i.e.  zero salt fraction the conductivity according to
our model is zero as it should ideally be. However  experimental 
observations indicate [3,4] that the
pure polymer has  non zero conductivities for reasons
 mentioned in ref.[2,17].
Our calculation could not be done for any other similar
complex due to lack of experimental data of
crystallinity (either from XRD or DTA). 

In
principle our model should be able to predict the optimum salt fraction
required to get the highest ionic conductivity for other complexes. The
difficulty remains, however, that the experimental crystallinity has been
used as input for different salt fractions, in our simulation. It is
necessary to predict the variation of crystallinity with salt fraction
from a consistent theory or at
least emperically to realise the full potential of our model.

The variation of crystallinity with X appears to be an interesting and
complex problem in itself, since available results show peculiar
irregularities and are often mutually contradicting[5]. We are at present
working on this problem and expect to report our findings later.
Some further shortcomings of our model are as follows. The effect
 of correlation of chain movements was not taken into
account. Also from optical micrographs and x-ray difraction analysis
 it is usually found that the crystalline regions are in
the form of spherullites [3]. Based on this, a better picture in
the model would have been  clustered groups of crystalline sites rather
than a random distribution of single crystalline sites. However, in view
of the dynamic disorder,  our model appears realistic, because a
single walker (i.e. a charge carrier) does not see  the whole structure,
but sees a small locality.

A useful extension of this model will be to incorporate variation of the
renewal time, the time in which the structure rearranges itself. In its 
present form the renewal time is the shortest possible i.e. one time unit
$\tau$. 
\begin{flushleft}
{\bf Acknowledgement :} \end{flushleft}

One of the authors (AJB) wishes to thank the UGC for providing financial
support in the form of a Senior Research Fellowship. The authors are also
grateful to Dr. T.K. Ballabh for useful suggestions regarding computer
 simulation.
\newpage {\begin{flushleft} {\bf References } \end{flushleft}
\begin{enumerate}
\item M.A. Ratner and D.F. Shriver, Chem. Rev., {\bf 88},109 (1988) and
references therein. 
\item C.A. Vincent, Prog. Solid State Chem., {\bf 17}, 145 (1987) and
references therein.
\item K.K. Maurya, N. Srivastava, S.A. Hashmi and S. Chandra,
 Jour. of Mater. Sci.,
    {\bf 27}, 6357 (1992).
\item N. Srivastava, A. Chandra and S. Chandra, Phys. Rev B
 {\bf 52}, 225 (1995).
\item D. Prusnikowska, W. Wieczorec, H. Wycislik, M. Siekerski and
J. Przyluski, Solid State Ionics, {\bf 72}, 152 (1994).
\item J.P. Bouchaud and A. Georges, Phys. Rep., {\bf 195}, 127 (1990).
\item S. Havlin and D. Ben-Avraham, Adv. in Phys., {\bf36}, 695 (1987).
\item J.W. Haus and K.W. Kehr, Phys. Rep., {\bf 150}, 263 (1987).
\item J.F. McCarthy, J. Phys. A, {\bf 26}, 315 (1986).
\item S.D. Druger, A. Nitzan and M.A. Ratner, J. Chem. Phys., {\bf 79},
 3133 (1983).
\item S.D. Druger, M.A. Ratner and A. Nitzan, Phys. Rev. B,
 {\bf 31}, 3939 (1985).
\item Aninda Jiban Bhattacharyya and S. Tarafdar, to be published.
\item J.B. Boyce and B.A. Huberman, Phys. Rep., {\bf 51}, 189 (1979).
\item S.H. Chung, K.R. Jeffrey and J.R. Stevens, J. Chem. Phys.,
 {\bf 93}, 1803 (1991).
\item C. Wang, Q. Liu, Q. Cao, Q. Meng and L. Yang,
 Solid State Ionics, {\bf53-56}, 1106 (1992).
\item W. Gorecki, R. Andreani, C. Berthier, M. Armand, M. Mali,
 J. Roos and D. Brinkmann,
     Solid State Ionics, {\bf 18-19}, 295 (1986).
\item S. Chandra, S.A. Hashmi and G. Prasad, Solid State Ionics,
 {\bf40-41}, 651 (1990).
\end{enumerate}
\newpage

{\bf Table-1 :Input parameters for calculation of theoretical
 conductivity and jump distance.}
\begin{center}
\begin{tabular}{|c|c|c|c|c|c|c|}
\hline \hline
Material & K & p$_{a}$ & p$_{c}$ & R=p$_{a}$/p$_{c}$ & $\tau(t_{a}+1)$ sec
 & D cm$^{2}$/sec \\
\hline \hline
PEO-NH$_{4}$SCN & 2.5 x $\times 10^{-4}$ & 0.999& 0.1&9.99  &
 7x10$^{-7}$ & 6x10$^{-10}$ \\
\hline
PEO-NH$_{4}$I & 1$\times 10^{-3}$ & 0.999& 0.1& 9.99 & 7x10$^{-7}$
 & 6x10$^{-10}$ \\
\hline\hline
\end{tabular}
\end{center}
\newpage
{\bf Figure captions : }\newline
1. Figure-1 Plot of theoretical conductivity versus salt fraction for
PEO-NH$_{4}$SCN. ($\Diamond$) show experimental results.[4]  \newline
2. Figure-2 Plot of theoretical conductivity versus salt fraction for
PEO-NH$_{4}$I. ($\Diamond$) show experimental results [3].
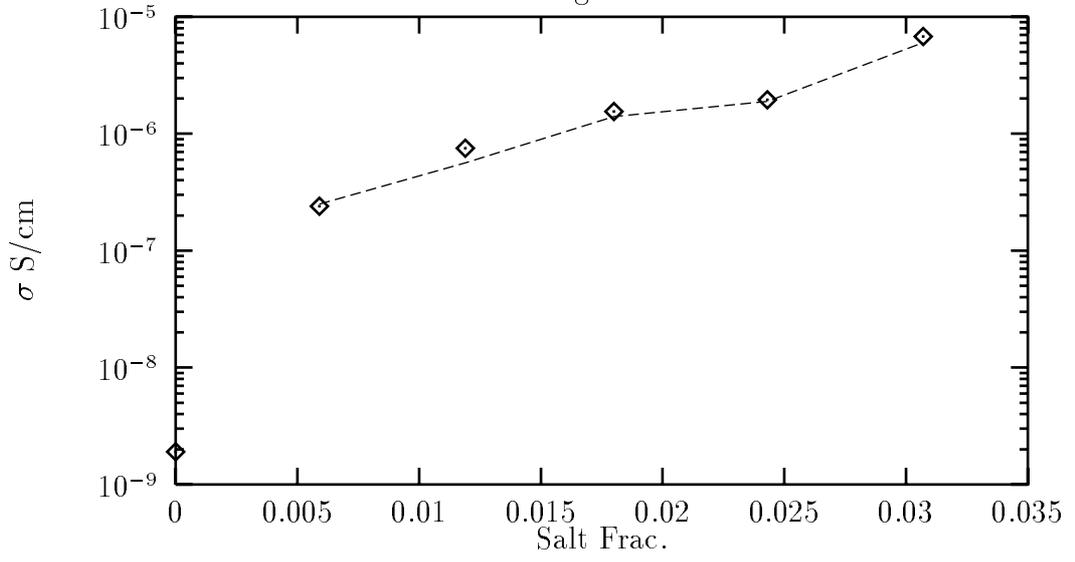
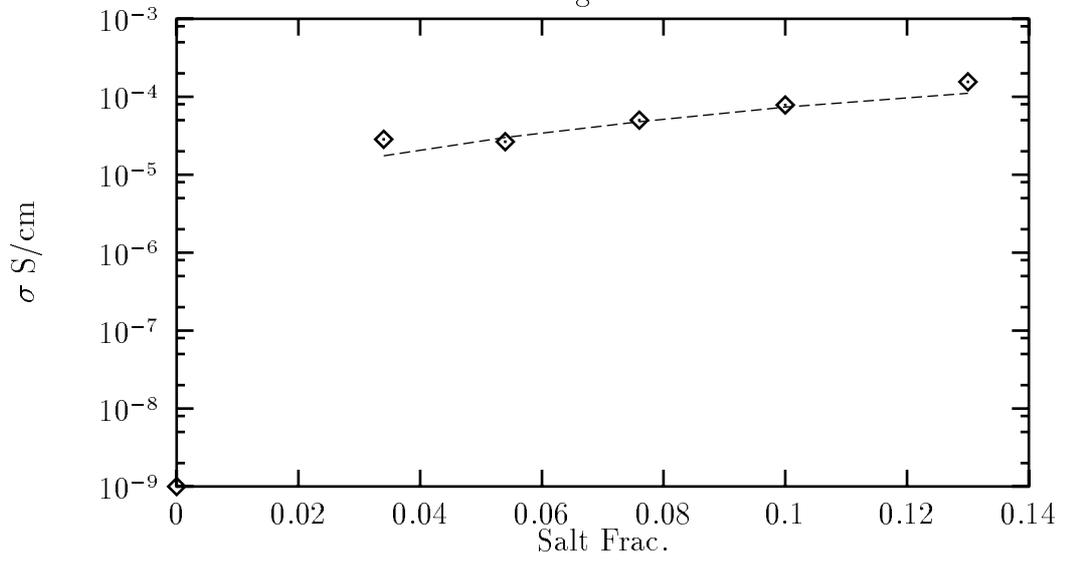
\end{document}